\documentclass{article}

\usepackage{PRIMEarxiv}

\usepackage[utf8]{inputenc} 
\usepackage[T1]{fontenc}    
\usepackage{hyperref}       
\usepackage{url}            
\usepackage{booktabs}       
\usepackage{amsfonts}       
\usepackage{nicefrac}       
\usepackage{microtype}      
\usepackage{lipsum}
\usepackage{fancyhdr}       
\usepackage{graphicx}       
\graphicspath{{media/}}     

\title{Correlated impact dynamics in science
}

\author{
  Jiazhen Liu \\
  Department of Physics \\
  University of Miami \\
  Coral Gables, Florida 33142, USA\\
   \And
  Tamang Kunal \\
  Department of Physics \\
  University of Miami \\
  Coral Gables, Florida 33142, USA\\
  \AND
  Dashun Wang \\
  Kellogg School of Management and McCormick School of Engineering \\
  Northwestern University \\
  Evanston, Illinois 60208, USA \\
  \And
  Chaoming Song \\
  Department of Physics \\
  University of Miami \\
  Coral Gables, Florida 33142, USA\\
  \texttt{c.song@miami.edu} \\
}

\begin{document}
\maketitle

\begin{abstract}
Science progresses by building upon previous discoveries. It is commonly believed that the impact of scientific papers, as measured by citations, is positively correlated with the impact of past discoveries built upon. However, analyzing over 30 million papers and nearly a billion citations across multiple disciplines, we find that there is a long-term positive citation correlation, but a negative short-term correlation. We demonstrate that the key to resolving this paradox lies in a new concept, called ``capacity", which captures the amount of originality remaining for a paper. We find there is an intimate link between capacity and impact dynamics that appears universal across the diverse fields we studied. The uncovered capacity measure not only explains the correlated impact dynamics across the sciences but also improves our understanding and predictions of high-impact discoveries. 
\end{abstract}


Isaac Newton's famous phrase, ``If I have seen further than others, it is by standing on the shoulders of giants'' \cite{newton2016if}, highlights the strong connection between a new discovery and the past advances that it builds upon. 
The correlation between scientific impact and previous knowledge was initially studied by Price \cite{price1965networks} through the construction of citation networks. Subsequently, qualitative research established that the impact of new scientific innovations is closely tied to the impact of prior knowledge \cite{foster2015tradition,fleming2001recombinant}. Recent large-scale empirical data analysis has statistically shown that papers with highly-cited references are more likely to receive high long-term citations \cite{mukherjee2017nearly,uzzi2013atypical}. In these studies, one of the most commonly used measures of impact is the number of citations a scientific work receives years after publication \cite{wang2013quantifying,lehmann2006measures,barabasi2012handful}.

On the other hand, recent development in network science provides a complementary insight into the impact of scientific works through the observation of the rich-club phenomenon in citation networks \cite{colizza2006detecting}. These studies suggest that new scientific papers connected to highly-cited older papers will have a greater impact themselves \cite{albert2002statistical,dorogovtsev2002evolution}. This phenomenon, known as degree assortativity, has been widely documented \cite{newman2002assortative}. Evidence supports the idea that the impact of prior literature plays a significant role in determining the impact of a paper, with those referencing highly-cited works more likely to receive high citations themselves \cite{mukherjee2017nearly}. Overall, the evidence supports Newton's hypothesis, suggesting that scientific works influenced by past "giants" are more likely to have a lasting impact.

Nonetheless, it has long been recognized that the long-term impact of references is not the sole determining factor for a paper's impact. Merton suggested that pioneering ideas are often inspired by recent information \cite{merton1961singletons}. This notion is supported by the observation that researchers only consider the immediate impact of inspiring works when pursuing new ideas, reflecting the current response of the community. Predicting the long-term impact of work requires foresight into the future development of the field. Many breakthrough discoveries are not immediately well received, a phenomenon referred to as the "Sleeping Beauty" effect \cite{van2004sleeping,redner2005citation,ke2015defining}. A recent study showed that scientific impact is strongly influenced by the promptness of innovation, consistent with Merton's theory \cite{mukherjee2017nearly}. However, existing studies verified a positive impact relationship between a paper and its references based on long-term citations. To the best of our knowledge, a direct test of Newton's hypothesis while conducting new research is still lacking in the literature. This raises an intriguing question: when a new innovation arises, what is the short-term impact correlation between the new innovation and old studies?

\section*{Results}

\begin{figure}[ht]
\centering
\includegraphics[width=1.0\textwidth]{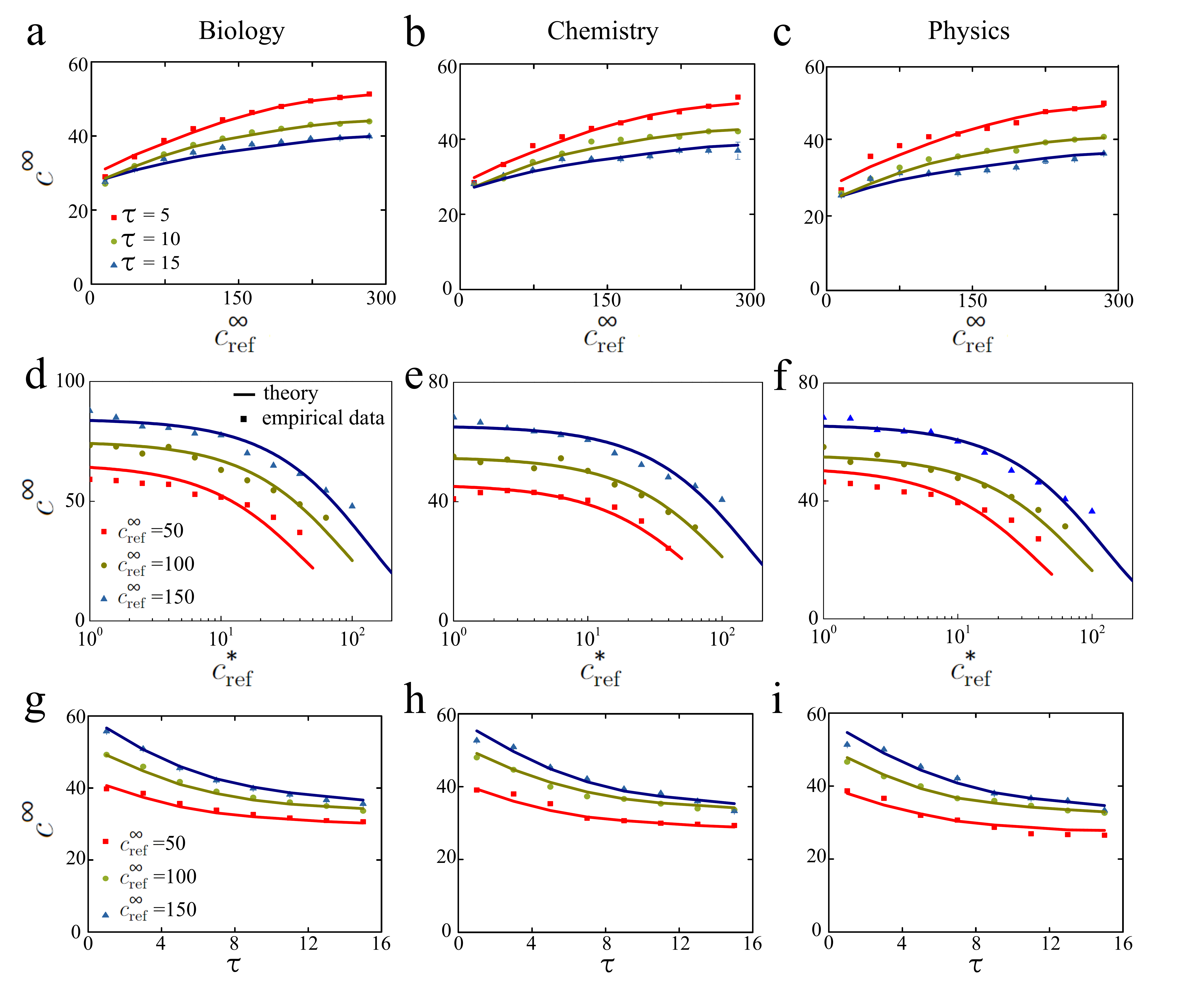}
\caption{\textbf{Empirical Measurements and Theoretical Prediction for Correlated Scientific Dynamics}. \textbf{a-c} show ${c^\infty}$ versus $c_{\rm{ref}}^\infty$ with fixed $\tau$ ($\tau=5yr,10yr$ and $15yr$) across biology, chemistry, and physics. \textbf{d-f} show ${c^\infty}$ versus $c_{\rm{ref}}^*$ with fixed $c_{\rm{ref}}^\infty$ ($\lambda=50,100$ and $150$). \textbf{g-i} show ${c^\infty}$ versus $\tau$ with fixed $c_{\rm{ref}}^\infty$ ($\lambda=50,100$ and $150$). The scatter plots are empirical observations whereas the solid lines are theoretical predictions. We find excellent agreement between theoretical predictions and empirical data across three different subjects.}
\label{fig:model}
\end{figure}

To answer this fundamental question, we collected data from the Web of Science (WOS) Thomson Reuters, which contains over 30 million papers and 830,259,176 citations from 1970 to 2014 (see Supplementary Material (SM) Data section). We perform two complementary tests. The first test was designed to investigate the relevance of a lagged citation-based metric used in previous studies, that is, the cumulative citations a paper receives on a long-term basis. To quantify the long-term impact, we measure the number of citations to a paper and its references 30 years after publication. This metric captures the established scientific impact of existing works from a retrospective perspective. Fig.~\ref{fig:model}a-c plots the long-term impact of the paper $c^\infty$ and its references $c_{\rm{ref}}^\infty$ in different fields of science, showing that the long-term impacts of papers and their references are positively correlated.

The second test was designed to explore the correlation of the impact when the new research was conducted. While our data do not indicate the exact implementation time, we use the publication time as a conservative measure. To quantify the immediate impact of a paper's references when the paper was just published, we measure its references' average citations $c_{\rm{ref}}^*$, reflecting the degree of innovation of this paper inspired by its related existing studies. To investigate the `short-term' citation correlation, by fixing the long-term impact $c_{\rm{ref}}^\infty$ of references, we measure the relationship between the long-term impact $c^\infty$ of a paper and the immediate impact $c_{\rm{ref}}^*$ of its references (Fig.~\ref{fig:model}d-f). If Newton's hypothesis were also valid in this case, one would expect $c_{\rm{ref}}^*$ to be positively correlated. Counterintuitively, we observe that $c^\infty$ decreases with $c_{\rm{ref}}^*$ (Fig.\ref{fig:model}d-f) across different domains and $c_{\rm{ref}}^\infty$. In particular, for high-cited papers ($c^\infty \geq 60$), $c_{\rm{ref}}^*$ is notably small ($c_{\rm{ref}}^* \leq 4$). This finding implies a rather surprising fact, most impactful innovations with large $c^\infty$ were built upon those ideas with a relatively small immediate impact $c_{\rm{ref}}^*$. 

The above two tests lead to seemingly contradictory results, creating a paradox of a long-term positive citation correlation but a simultaneous short-term negative correlation. To understand this paradox better, we demonstrate a well-known example in condensed state physics \cite{redner2005citation}. In 1955, Goodenough developed a fundamental theory to predict magnetism in transition-metal oxides \cite{goodenough1955theory}. This work did not attract much attention from the condensed stated physics field until the 1990s. Due to the development of experimental techniques, several studies \cite{radaelli1997charge, radaelli1995simultaneous,mitchell1996structural} began to focus on Goodenough's extraordinary work. At the time of this experimental work (the mid-1990s), Goodenough's work had fewer than 15 citations, i.e., $c_{\rm{ref}}^*<15$. Both Goodenough's papers and those of his followers ended up being highly cited and had a huge impact on physics and materials science. This example explains the observed paradox: these experimental works, compared with the thousands of other follow-up papers that cite Goodenough's work, recognize the importance of Goodenough's predictions at a time when the quest for high-temperature superconductivity is still in its very early stages. This observation is consistent with the finding in Fig.~\ref{fig:model}a-f: the breakthrough is rooted in its insights of foreseeing a fruitful innovation inspired by a future-impactful (large $c_{\rm{ref}}^\infty$) but yet-to-be-recognized discovery (small $c_{\rm{ref}}^*$), i.e. future giants. This proposes a new paradigm of scientific impact correlation that differs from the original Newton's hypothesis. 

To investigate this new paradigm quantitatively, we introduce a novel metric that captures both positive long-term and negative short-term correlations simultaneously. We define a paper's $capacity$
\begin{equation}\label{eq:phi_def}
\phi=\frac{\langle\Delta c_{\rm{ref}}\rangle}{\langle c_{\rm{ref}}^\infty\rangle}=1-\frac{\langle c_{\rm{ref}}^*\rangle}{\langle c_{\rm{ref}}^\infty\rangle},
\end{equation}
being the normalized $\Delta c_{\rm{ref}}$ averaged all references. 
Intuitively, the capacity $\phi$ quantifies the remaining fraction of originality for the paper. By definition, $\phi$ increases with $c_{\rm{ref}}^\infty$ whereas decreases with $c_{\rm{ref}}^*$, in line with the observation in Fig.~\ref{fig:model}. It is worth pointing out that the capacity $\phi$ is exclusive. When a paper acquires a large capacity $\phi$ from an existing idea, the subsequent followers of this prior work can only achieve smaller ones, implying underlying competition among new papers that are inspired by the same existing ideas. We will show below that the $capacity$ $\phi$ encodes all information about correlated impact dynamics.

\begin{figure}[ht]
\centering
\includegraphics[width=1.0\textwidth]{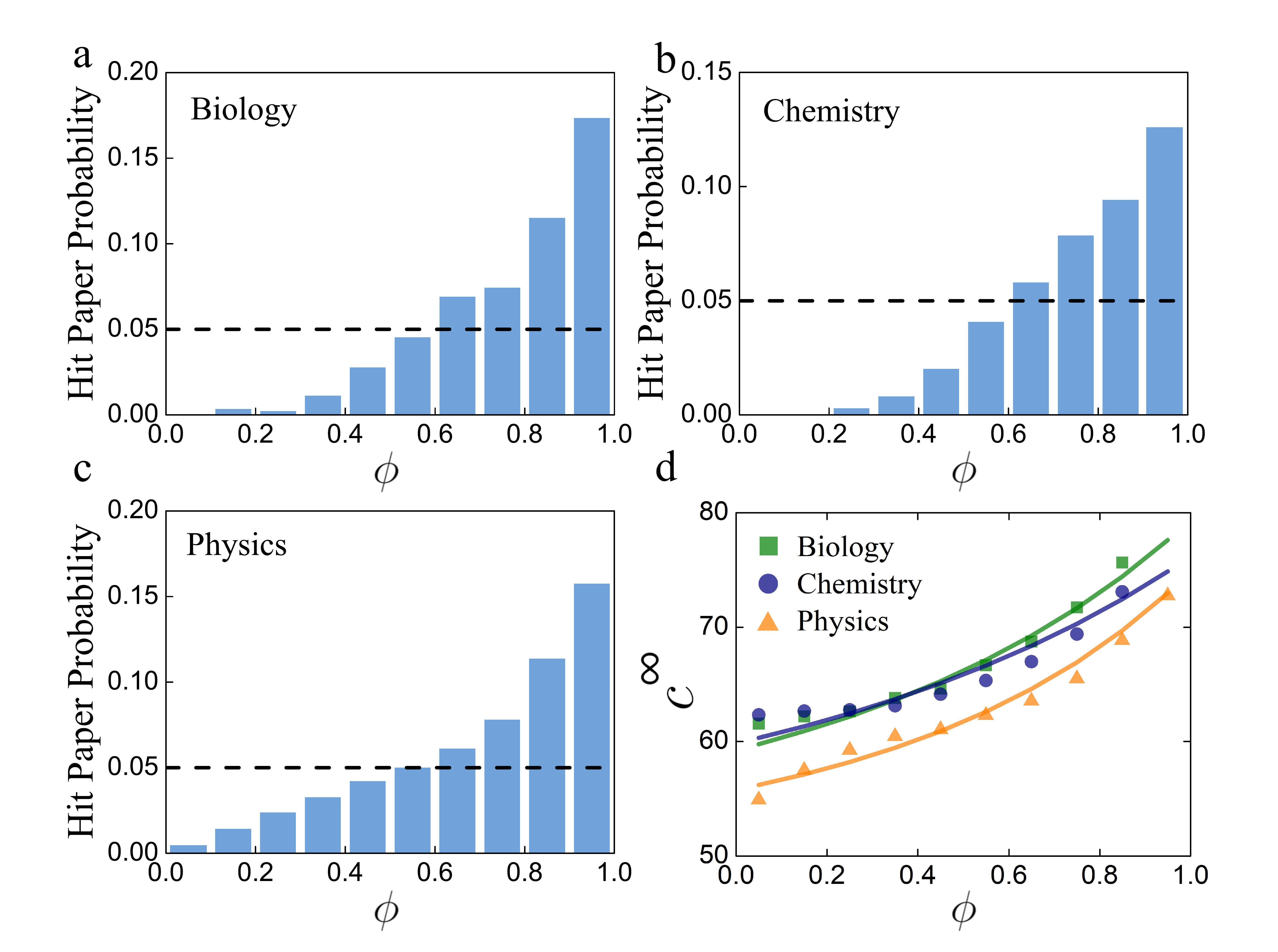}
\caption{\textbf{Hit Paper Probability \& Impact vs Capacity}. By using the published in Web of Science from 1970-2014, \textbf{a-c} measure the probability of a paper being in the top 5\% of the $c^{30}$ distribution with the capacity $\phi$ across biology (\textbf{a}), chemistry (\textbf{b}), and physics (\textbf{c}). Papers with capacity $\phi\geq0.9$ have a hit rate of about 15\%, which is about triple the background rate of 5\%. Papers with capacity $\phi\leq0.6$ have a hit rate of about 4\%, less than the background rate of 5\%. This finding suggests that papers with high capacity have a higher chance of becoming impactful papers. The supplementary materials present highly similar findings when considering 'hit' papers defined as the top 1 \% or 10 \% by citations, hinting at the universality of the discovered correlation between papers' impact and capacity. \textbf{d} measure the average long-term impact $c^\infty$ with the capacity $\phi$ across biology (green), chemistry (blue), and physics (orange). We discover that the impacts of papers grow exponentially with capacity, where the slopes of the dash lines $\alpha$ capture the strength of the correlation, equal to $0.98$, $1.00$, and $1.62$ for biology, chemistry, and physics respectively. }
\label{fig:capacity}
\end{figure}

To explore the relationship between a paper's long-term impact and its capacity received from its references, we study ``hit'' papers whose citations are in the top 5 percentile in the dataset \cite{uzzi2013atypical}. Figure \ref{fig:capacity}a-c presents the probability of finding 'hit' papers with capacity $\phi$ across four different subjects. We discover that the probability of finding a 'hit' paper increases rapidly with capacity $\phi$ across all the subjects. Papers with a large capacity, i.e., $0.9\leq\phi\leq1$, display a hit rate of around $15$ out of $100$ papers, which is about triple the background rate of $5$ out of $100$. On the contrary, when the capacity $\phi$ of a paper is relatively small, we find significantly lower hit rates across all the subjects. Papers with capacity $\phi<0.6$ show hit rates of around $4$ out of $100$ papers, lower than the background rate. Thus, our findings indicate that a paper with a large capacity is more likely to be impactful in the future. Indeed, based on the definition of capacity, a paper with higher capacity acquires more originality and novelty from the prior ideas, leading to greater importance and higher future impact. We further consider different definitions of impactful scientific works (see SM S2 Empirical Results), i.e., ``hit'' papers, finding the same patterns we observe in Fig.\ref{fig:capacity}a-c.

We further measure the long-term impact of papers as a function of capacity. Fig.\ref{fig:capacity}d shows the average long-term impact increase double-exponentially with capacity across three different subjects, satisfying
\begin{equation}\label{exponential}
    \ln c^\infty =  A e^{\alpha\phi} + B ,
\end{equation}
where parameters $A$, $B$, and $\alpha$ are constants depending only on the fields. In particular, the scaling factor $\alpha$ captures the slope of the solid line in figure \ref{fig:capacity}d , indicating a measure of the strength of the correlation between long-term impact and capacity. Comparing the $\alpha$ of three different subjects, we find the $\alpha$ of physics has the largest correlation $\alpha=1.6$, indicating the impact of physics papers relies strongly on existing works, whereas biology and chemistry have a relatively weaker correlation with $\alpha\approx 1$ (Figure \ref{fig:capacity}d). The discoveries of Fig.\ref{fig:capacity} suggest the universality of the correlation between scientific impact and capacity, implying that the long-term impact of new papers can be predicted by capacity $\phi$. We next show that a model based on the discovered correlation naturally leads to the empirically observed scientific impact. 

A current study \cite{wang2013quantifying} has discovered that the time evolution of the cumulative citation, $c^t$, the number of citations the paper acquires after $t$ years publication, follows a unique function $c^t=m[e^{\lambda \Phi(\frac{\ln {t}-\mu}{\sigma})}-1]$, 
where $\Phi(x)=(2\pi)^{-1/2}\int_{-\infty}^{x}e^{-y^2/2}\mathrm{d}y$. While $\mu$ and $\sigma$ control the shaping of $c^t$, the long-term impacts of papers also depend on the fitness parameter $\lambda$ and average citation $m$. Incorporating Eq.~(\ref{eq:phi_def}) and ~(\ref{exponential}) with $c^t$ allows us to predict papers' long-term impacts $c^\infty$ with their references' long-term impacts $c_{\rm{ref}}^\infty$ and instant impact $c_{\rm{ref}}^*$. Indeed, as Fig.~\ref{fig:model}a-f demonstrates, our theoretical predictions (solid lines) precisely agree with the empirical observations (scatters), indicating that the correlated impact dynamic is fully captured by the capacity $\phi$.

The capacity also provides a comprehensive explanation of the previous observation on the correlation between scientific impacts and publication immediacy $\tau=t-t_{ref}$, capturing the published time difference between a paper and its references \cite{mukherjee2017nearly}. We plot $c^\infty$ as a function of publication immediacy $\tau$ (Fig.~\ref{fig:model}h-i). By accelerating publication immediacy, we observe a downward trend in the long-term impact of papers, similar to Fig.\ref{fig:model}d-f. Indeed, as the number of citations received by a paper grows with time, the immediate impact $c_{\rm{ref}}^*$ also increases with publication immediacy $\tau$. In other words, the $c_{\rm{ref}}^*$ not only measures the immediate impact of references but also accounts for the fading novelty of the past scientific literature. The theoretical prediction further confirms the above statement, finding perfect matches between the prediction and empirical data. Generally, the capacity $\phi$ accounts for the correlated impact dynamics by correlating a paper's long-term impact with the fading novelty and ultimate impact of its references. 

\begin{figure}[ht]
\centering
\includegraphics[width=1.0\textwidth]{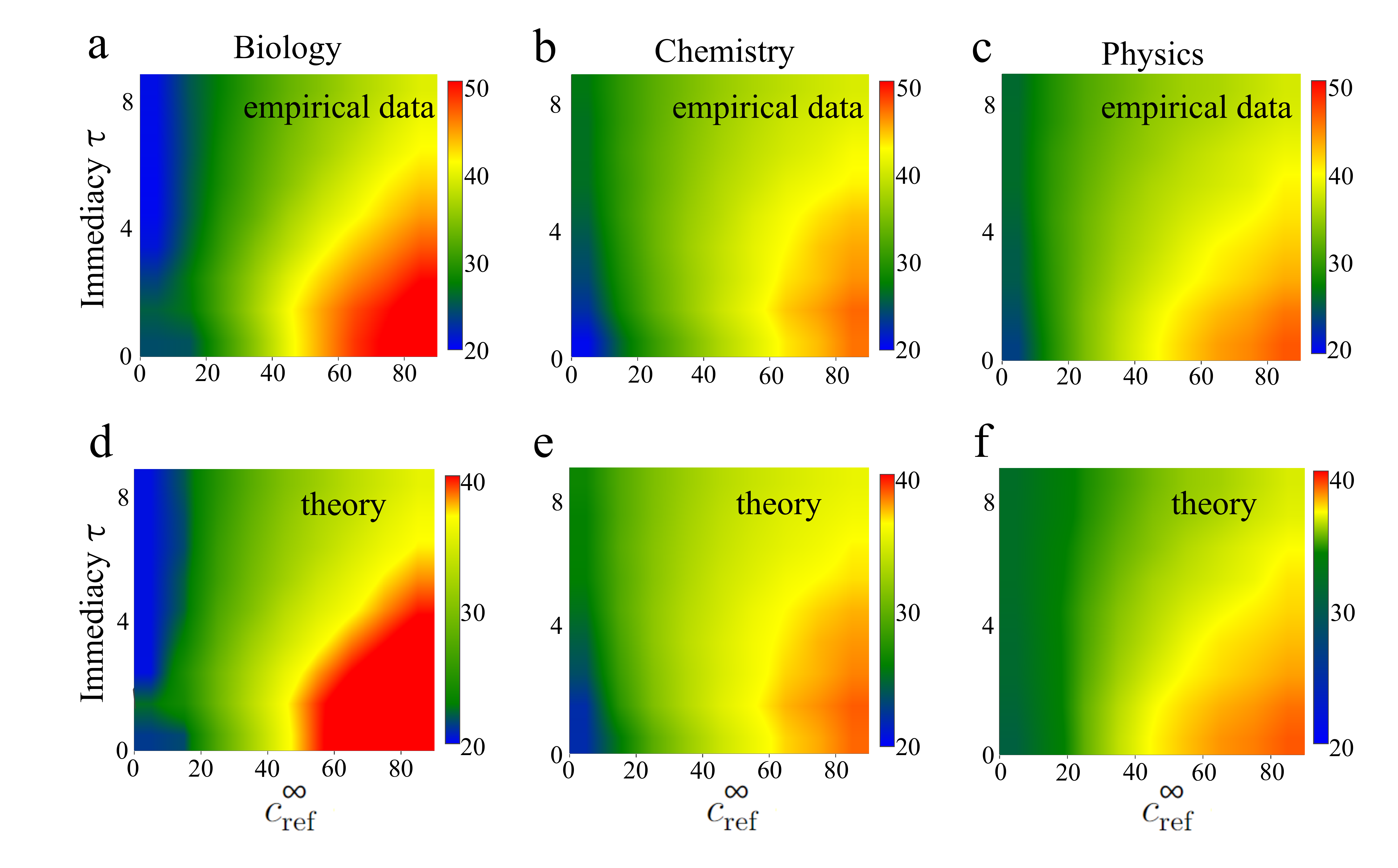}
\caption{\textbf{Correlation Heatmaps}. Correlation of reference impact ($c_{\rm{ref}}^\infty$) and publication immediacy ($\tau$) with capacity $\phi$, where (\textbf{a-c}) show empirical data and (\textbf{d-f}) present theoretical predictions across biology (\textbf{a, d}), chemistry (\textbf{b, e}), and physics (\textbf{c,f}). Figures show highly similar patterns between empirical observations and modeling results, implying the impact correlation is governed by the proposed metric $capacity$.}
\label{fig:exp}
\end{figure}

The excellent agreement between theoretical predictions and empirical observations implies that the capacity $\phi$ defined in Eq.~(\ref{eq:phi_def}) plays an important role in predicting the impact correlation between papers and their references. To further validate the proposed model, we investigate the correlation by plotting the papers' impact $c^\infty$ as a function of their references' long-term impact $c^\infty_{\rm{ref}}$ and publication immediacy $\tau$ for empirical data (Fig.~\ref{fig:exp}a-c). Moreover, our model allows us to predict the correlation of impact between papers and their references (Fig.~\ref{fig:exp}d-f), finding surprising agreements between the empirical measurements and modeling predictions. These strikingly similar patterns suggest that a paper's long-term impact is strongly correlated with the prior works by the capacity $\phi$ (Eq.(\ref{eq:phi_def})). 

Inspired by the success of capturing the correlated impacts of papers and their references, one may wonder whether the capacity can be used to discover the breakthrough papers, i.e., giants. The breakthrough papers, characterized by groundbreaking achievements and unparalleled profound impacts, are considered to be different from the ordinary impactful papers \cite{li2019dataset}. A long-standing problem of the science of science is to understand and discover breakthrough papers, e.g., Nobel-Prize-winning papers, in their early stage \cite{zuckerman1967nobel,fortunato2014growing}. Our findings offer a potential solution. In Fig.~\ref{fig:exp}, both empirical observations and theoretical predictions suggest that the most influential papers appear in the same region, characterized by significant reference long-term impact $c^\infty_{\rm{ref}}$ and small publication immediacy $\tau$, results in a large capacity $\phi$ based on Eq.~\ref{eq:phi_def}. Considering the extremely far-reaching impacts of breakthrough papers, this finding suggests that the most groundbreaking scientific works are probably characterized by significantly large capacity. 

\begin{figure}[ht]
\centering
\includegraphics[width=1.0\textwidth]{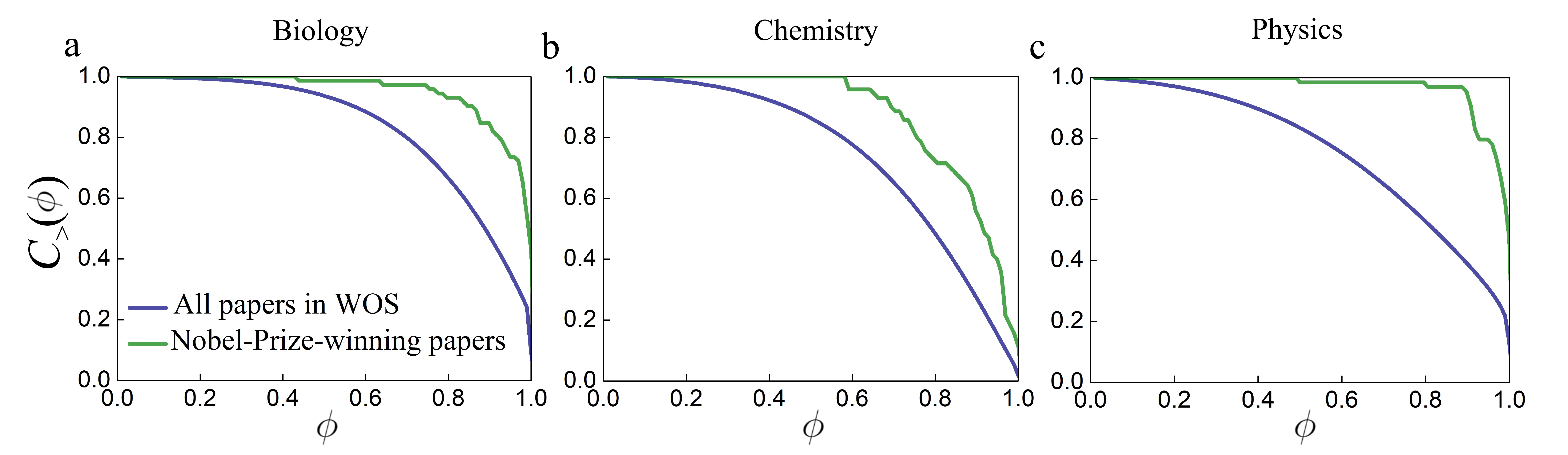}
\caption{\textbf{The Capacity of the Breakthrough Papers}.  In order to compare with the papers in WOS, we select the Nobel Prize-winning papers published after 1970 for our measurement. We collect a total of 74,73, and 64 prize-winning papers in biology, chemistry, and physics. \textbf{a-c} measure the complementary cumulative distribution (CCDF) of capacity, $C_>(\phi)$, across different subjects. The green plots measure the $C_>(\phi)$ for the Nobel-Prize-winning papers, whereas the blue plots measure the $C_>(\phi)$ for all papers in the field. These plots suggest that the Nobel Prize-winning papers have a significantly larger capacity than other papers.}
\label{fig:breakthrough}
\end{figure}

To further confirm our thought, we collect $74$, $73$, and $64$ Nobel-Prize-winning papers published between 1970-2014 as the representatives of breakthrough papers \cite{li2019dataset} for biology, chemistry, and physics, respectively. Figure \ref{fig:breakthrough}a-c plots the complementary cumulative distribution function $CCDF$ of capacity $\phi$, $C_>(\phi)$, that measures the proportion of papers with its capacity larger than $\phi$. We find that the Nobel-Prize-winning papers have a significantly larger capacity $\phi$ than normal papers. These findings imply that breakthrough papers with groundbreaking impact are determined mainly by their capacity $\phi$. Indeed, breakthrough research not only has a broad impact, it often breaks new ground, i.e., the pioneering work. Hence, it is required to recognize the area of great potential for innovation at a very early stage, leading to an extremely high capacity. 


\section*{Discussion}

Our study challenges the common belief that there is a positive relationship between new discoveries and the impact of existing works on which they are based. In contrast, we found a negative correlation with immediate impact, that is, the impact of references when a paper is published, despite a positive correlation between long-term impact. This is because the innovation has a limited amount of capacity, and fades off after more new works recognize and further develop these new ideas. Therefore, it implies that not only the significance of the underlying ideas but also the promptness of recognizing their importance are equally important for subsequent innovations. We show that ``capacity'', measuring the proportion of remaining originality in existing ideas, encapsulates all information about correlated impact dynamics and captures both positive long-term and negative short-term citation correlations. We discovered a universal relationship between a paper's impact and its references' capacity, and develop a theory that accurately predicts a paper's impact. Our model provides a generic mechanism for understanding the emergence of degree correlations in complex networks and sheds new light on complex network dynamics.



Furthermore, our findings have implications for identifying scientific breakthroughs and discovering potentially groundbreaking papers at an early stage. It has been challenging to differentiate breakthrough papers from ordinary impactful papers using only citations. For instance, a highly-cited paper might receive a large number of citations, but it might not be a pioneering work in its field. Hence, citations alone are insufficient in characterizing breakthrough papers. Our proposed metric, $capacity$, may serve as a potential solution to differentiate between breakthrough papers and regular papers. Our results show that Nobel Prize-winning papers, as examples of breakthrough works, have significantly higher capacity compared to regular papers. This indicates that groundbreaking innovations are characterized by extremely high $capacity$.

On the other hand, our study only considers the correlated impact dynamics on paper citation networks. Previous research has shown that the size of a research team has an influence on the likelihood of a breakthrough discovery \cite{wu2019large}. Additionally, the impact and long-term citation of a paper have been found to be related to the author's reputation and productivity \cite{petersen2014reputation,sinatra2016quantifying}. There is evidence of author ``hot streaks" where a high concentration of impactful papers are produced around a specific topic \cite{liu2018hot}. Further research is needed to explore the combined effect of our proposed metric $capacity$ and these relevant factors on breakthrough innovations.

\section*{Method}
\subsection*{Data Description And Processing}
\subsubsection*{Web Of Science Data}
We use the Web of Science (WOS) dataset from 1970-2014. It consists of 43,661,391 publications and 800M citations among them. Figure S1 plots the number of papers with time, showing exponential growth in line with the previous findings \cite{price1965networks} (see Supplementary Material) .

To classify these papers into Biology, Chemistry, Math, and Physics subjects, we start with four sets of papers based on the WOS journal categories of keywords `bio-’, `chem-’, `math’, and `physics’, each forming a core of the corresponding subject. For each subject, we further include all papers that either cite or are cited by any papers in the core. In the end, we obtain four citation networks with 13M, 9M, 3M, and 7M nodes for biology, chemistry, math, and physics, respectively. Table S1 summarizes the basic statistics of these four networks, each being a sub-graph of the whole WOS citation network (see Supplementary Material). Note that there are overlaps among these sub-graphs because of multidisciplinary papers. Figure S2 plots the Venn diagram of four networks, annotating interdisciplinary overlaps (see Supplementary Material). We find there are large overlaps between Biology \& Chemistry and Chemistry \& Physics, whereas Math has the smallest proportion of papers that belong to multi-disciplines.

\subsubsection*{Nobel-Prize-winning Papers}
We use the Nobel Prize-winning dataset collected by Li et.al. \cite{li2019dataset}. The dataset contains 230, 277, and 236 papers awarded the Nobel Prize between 1901-2016 in Chemistry, Medicine, and Physics, respectively. During 1970-2014, the dataset consisted of 77, 77, and 64 papers for each category. The WOS dataset covers the vast majority of these papers, containing 74, 73, and 64 papers, respectively. Table S2 summarizes a breakdown of these Nobel-Prize-Winning papers into four subjects and multidisciplinary subjects created in Section S1 (see Supplementary Material). We find that a large overlap exists between Biology and Chemistry. 


\bibliographystyle{unsrt}  
\bibliography{references}  
\newpage

\newpage

\newpage

\newpage

\newpage

\end{document}